# Local laser oxidation of titanium film for post-fabrication trimming of photonic integrated circuits

**Aleksandr V. Tronev**[1,*], **Mikhail V. Parfenov**[1], **Sergey I. Bozhko**[2], **Andrey M. Ionov**[2], **Rais N. Mozhchil**[2], **Sergey V. Chekmazov**[2], **Petr M. Agruzov**[1], **Igor V. Ilichev**[1], **Aleksandr V. Shamrai**[1]

[1]*Laboratory of Quantum Electronics, Ioffe Institute, Saint Petersburg, 194021, Russian Federation*
[2]*Laboratory of Semiconductor Surfaces Spectroscopy, Osipyan Institute of Solid State Physics Russian Academy of Science, Chernogolovka, 142432, Russian Federation*
*\* a.tronev@mail.ioffe.ru*

**Abstract:** Local laser oxidation of a thin titanium film is considered as a means of a precise adjustment of losses and effective refractive index of dielectric optical waveguides. A fine phase control of an operating point and extinction ratio enhancement up to 57 dB were demonstrated using an integrated optical Ti:LiNbO$_3$ Mach-Zehnder modulator. This technique only slightly affects the dielectric waveguide material and is very promising for a high precision permanent trimming of photonic devices based on dielectric waveguides of different material platforms and fabrication technologies.

## 1. Introduction

Photonic integrated circuits (PICs) have attracted a great attention from both the academic community and industry. In particular, dielectric waveguide based on photonic devices have experienced tremendous progress in recent years and show a strong uptrend in various applications such as telecommunications [1], microwave photonics [2], optical sensing [3] and quantum information technologies [4]. Whereas silicon-based photonics is considered the most promising [5] owing to low waveguide loss and compatibility with CMOS fabrication processes, other material platforms for fabricating dielectric waveguide as well being developed to further extend beyond what silicon has to offer. They include lithium niobate [6], nitrides [7], diamond [8], tantalum pentoxide [9], and some polymers [10].

All the material platforms mentioned above have similar fabrication challenges related to local variations in the waveguide dimensions and the refractive index due to technological imperfections. This results in deviations of the device performance, which will differ slightly from its ideal design at a wafer-scale level. It is common practice to actively tune PICs elements by employing electro-optic [11] and thermo-optic [12] effects, free carrier dispersion [13], or other principles depending on the material platform. This solution has some disadvantages, including related with power consumption, sophisticated control electronics, and crosstalk.

An alternative solution is a permanent modification of the waveguide structure, a tailoring of the guided mode properties that counterbalances the fabrication imperfections to achieve the desired device characteristics. This post-fabrication trimming (PFT) technique can provide a much simpler set-it-once-and-forget-it mechanism for adjusting photonic components, which is typically achieved by permanently changing the effective refractive index of optical waveguides and, hence, does not require a constant supply of power. Oxide materials are widely used in different PFT techniques.

One can divide PTF techniques into two groups: modification of waveguide core and trimming of the waveguide cladding. Laser ablation [14] illustrates the type of PFT via the core modification. This is a universal technique that can be used for any material platforms; however, it is rather a destructive method with damage to the waveguide material and structure. Other techniques for a waveguide core modification such as local amorphization with nanomilling via femtosecond laser irradiation [15], localized annealing of an ion-implanted waveguide section [16,17], UV irradiation of a hydrogenated amorphous silicon waveguide [18], and local oxidation with an atomic force microscope probe [19] were developed for silicon photonics and cannot be directly applied to other substrate materials.

An additional cover layer is modified in trimming of the waveguide cladding. This approach does not deteriorate the waveguide core, it is more universal and can be used for many material platforms [20–24]. Oxide materials are a good choice for an additional cover layer. For example, trimming via induced compaction and strain of silicon dioxide ($SiO_2$) has been demonstrated. Indirect application of mechanical stress to the waveguide core is necessity due to rather low refractive index of silicon dioxide. This technique suffers from the long-term stability problem, since the level of $SiO_2$ cladding compaction relaxes over time [22].

Titanium dioxide ($TiO_2$) has one of the highest refractive index among oxide materials [25] and a rather low optical absorption in a wide wavelength range, so that it could be used for PFT via penetration directional light into the cover layer without any effect on the waveguide core.

In previous works [26,27], we demonstrated the modification of a 5 nm thick titanium film on the top of a titanium in-diffused lithium niobate (Ti:$LiNbO_3$) waveguide by laser irradiation and supposed that this technique can be applied to a precise adjustment of the characteristics of integrated optical circuits.

In this paper, we perform our theoretical analysis of the influence of titanium and titanium dioxide top layers on optical waveguides with different refractive index contrast (numerical aperture) carried out for verifying applicability of the technique for PFT of PIC. We have explored the mechanism of local laser oxidation in more detail to show that this technique is compatible with dielectric waveguide of various material platforms. As an experimental demonstration, we applied the proposed technique to improve the performance of a Mach-Zehnder modulator based on Ti:$LiNbO_3$ waveguides. This material platform is very convenient, since a large size of the waveguides and transparent substrate simplify the technical implementation. A fine phase control with a step of 0.013 rad and an increase in the extinction ratio (ER) by 27 dB (up to 57 dB) were demonstrated with real time monitoring of the modulator transmission characteristic. No long-term trimming relaxation was observed during several months of a periodical monitoring. These results may be interesting for some applications such as quantum telecommunications and precise optical measurements. Note that the change in the effective refractive index of the waveguide will be higher for the waveguides with a higher refractive index contrast and a smaller mode field diameter (MFD), such as those fabricated on the basis of thin-film lithium niobate (TFLN), silicon nitride, or silicon on insulator (SOI). Thus, more efficient photonic device trimming in a wider range of parameters is expected for these material platforms. However, more precise laser beam control, focusing and alignment will be required.

## 2. Thin films on the top of dielectric optical waveguides

The physical principals of the influence of a cover thin film on an optical waveguide can be explained by a simplified model of zigzag beam propagation in a four-layer waveguide structure (Fig 1) [26].

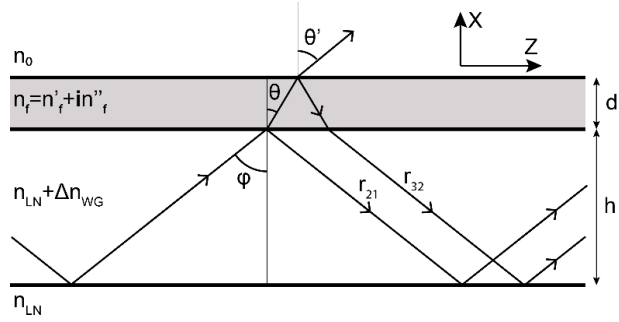
Fig. 1. Zigzag beam propagation in a four-layer waveguide structure.

The change in the waveguide effective refractive index and the addition optical loss per unit length due to the thin film can be estimated in terms of the film reflection coefficient [28]:

$$\Delta n_{eff} = \frac{\lambda}{2\pi} \cdot \frac{\phi_f}{2h \cdot \tan(\varphi)}, \quad \Delta\alpha = \frac{1-R_f}{2w \cdot \tan(\varphi)}, \quad (1)$$

where $\lambda$ is the light wavelength, $h$ is the waveguide height, $\varphi$ is the angle of beam propagation in the waveguide mode (a single mode waveguide is considered), $R_f$ and $\phi_f$ are the amplitude and phase of the film complex reflection coefficient $r_f$, respectively. Note that in this simplified model, an additional optical loss is a result of combination of film absorption and leakage due to violation of total internal reflection ($R_f < 1$). The film reflection results from the interference of reflection from waveguide/film and film/air interferences

$$r_f = \sqrt{R_f}\exp(i\phi_f) = \frac{r_{32} + r_{21}\exp(i\delta)}{1 + r_{32}r_{21}\exp(i\delta)}, \quad \delta = 2\frac{2\pi}{\lambda}n_f d \cos\theta. \quad (2)$$

Here $r_{32}$ and $r_{21}$ are the Fresnel reflection coefficients for the waveguide-film and film-air interface respectively, the phase shift $\delta$ depends on the film thickness $d$ and the film refractive index $n_f = n'_f + in''_f$, which is a complex value in the general case. Qualitative analysis shows that the higher the film refractive index, the greater its influence on the optical waveguide. Note that the film thickness is limited by the cutoff conditions of higher order modes.

A titanium metal film is a promising candidate for fine tuning of waveguide properties. It is characterized by a high real part of refractive index $n'_{Ti} = 3.68$ at 1550 nm [29], which even exceeds the refractive index of silicon. Thus, it can be used for most of material platforms for dielectric waveguide fabrication. A moderate imaginary part of refractive index $n''_{Ti} = 4.53$ [29] gives an acceptable excess optical loss for a properly chosen film thickness. Different influences on the TE and TM polarization modes of the waveguide due to plasmon polariton excitation [30] give a possibility of the light polarization control.

Titanium (Ti) is quite attractive from the technological aspect due to its compatibility with CMOS. DC magnetron sputtering can be used for a high-quality deposition of films with stable optical characteristics, which have a good adhesion to most materials of dielectric waveguides. A thin Ti film can be locally oxidized by laser illumination [31–34], which will be considered in detail in the next section. $TiO_2$ is a well-known material in integrated optics [35]. For example, it is used for hybrid waveguides on thin film lithium niobate [36,37] and as a corrective layer for thermal configuration of silicon photonic devices [38]. This is also a very stable material with significantly different optical properties as compared with Ti. Thus, laser oxidation provides a way for a permanent laser trimming of PICs with a little additional space consumption.

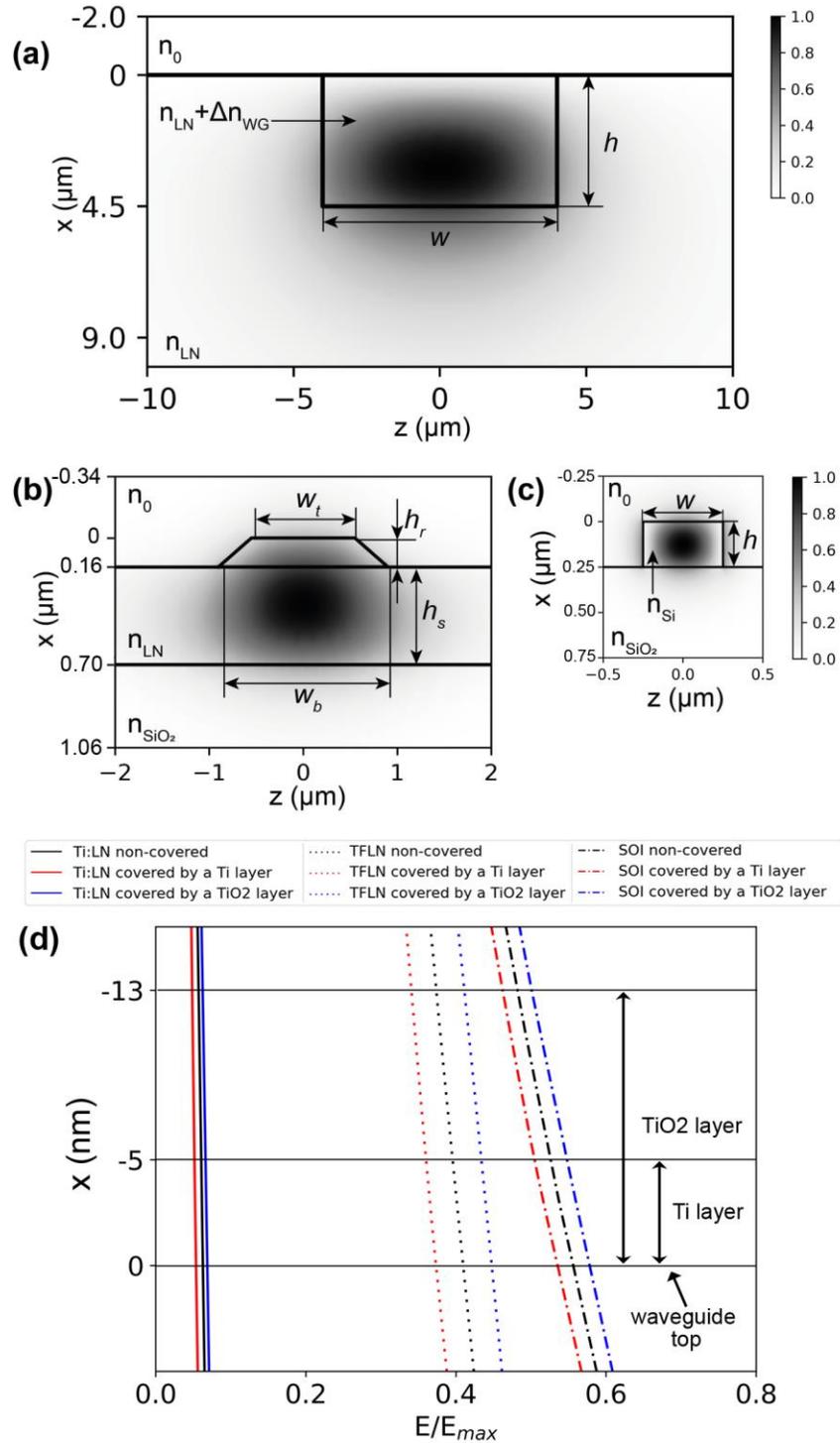

Fig. 2. Channel waveguide and calculated electric field distributions of TE modes used in the simulation for: (a) Ti-indiffused lithium niobate (Ti:LiNbO$_3$) waveguide; (b) thin-film lithium niobate (TFLN) rib waveguide; (c) silicon-on-insulator (SOI) waveguide; (d) dependence of normalized electric field distribution near the upper boundary of the waveguide without film, for metallic Ti film and for dielectric TiO$_2$ film.

A quantitative estimation of thin Ti and TiO$_2$ films effect on dielectric optical waveguides was carried out using numerical simulation by the finite element method. Three material platforms for optical waveguide fabrication were considered. These were a Ti indiffused lithium niobate (Ti:LiNbO$_3$) waveguide [39], a thin-film lithium niobate (TFLN) waveguide in the form of an etched rib [40], and a silicon-on-insulator (SOI) waveguide [41]. The material platforms chosen represent three levels of refractive index contrast and MFD. In the case of Ti:LiNbO$_3$ waveguides, the refractive index contrast (~ 10$^{-3}$) is low, and the mode is loosely confined (MFD ~ 10 µm). TFLN waveguides have a high refractive index contrast (~ 1) and a tight mode confinement (MFD 1-2 µm). The SOI waveguides have the highest refractive index difference (~ 2) and the lowest MFD (< 1 µm). Other material platforms can be subsumed to one of these three levels of refractive index contrast and MFD.

We limit ourselves to analysis of a horizontally polarized (TE) waveguide mode, which is used in electro-optic modulators. Note that Ti film influence on orthogonal TM mode can be higher due to plasmon-polariton mode excitation [30]. However, the light propagation conditions for very thin films (several nm thick), which, as shown below, are preferred for laser oxidation, are far from the surface plasmon-polariton mode excitation [42], and the Ti film influence on the TM mode is similar to effect on the TE mode.

Typical parameters of single-mode waveguides at wavelength of 1550 nm were used in the analysis for each material platform and were presented earlier in experimental works [43–47]. To reduce the computational complexity, the Ti:LiNbO$_3$ gradient waveguide was replaced by step-index waveguides with a width $w = 8$ µm, a height $h=4.5$ µm and a refractive index difference $\Delta n = 5 \times 10^{-3}$ [43] (Fig. 2a). The TFLN waveguide rib had a width $w_t = 1.1$ µm at the top and a width $w_b = 1.8$ µm at its base and a height $h_r = 160$ nm [44] (Fig. 2b). The slab height $h_s = 540$ nm. For the waveguides directed along the $y$ crystal axis on an $x$-cut LiNbO$_3$ single crystal substrate, the TE mode was an extraordinary wave. Thus, the extraordinary refractive index $n_e = 2.14$ was used. The SOI waveguide channel had the smallest sizes $w = 500$ nm and a height $h = 250$ nm [45] (Fig. 2c). The refractive indexes of silicon dioxide and silicon at wavelength of 1550 nm were set as $n_{SiO2} = 1.44$ [46] and $n_{Si} = 3.48$ [47].

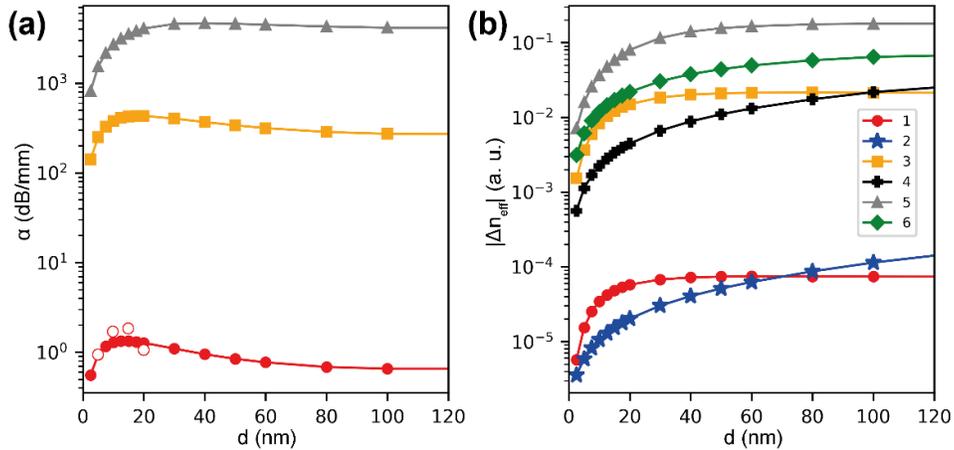

Fig. 3. Propagation losses (a) and refractive index differences (b) of the modes induced by the presence of Ti and TiO$_2$ layers of different thicknesses d on the waveguide top. 1 (circles) – Ti:LiNbO$_3$ with Ti layer (results of measurements of waveguide loss change per unit length at laser oxidation are shown as open circles); 2 (stars) – Ti:LiNbO$_3$ with TiO$_2$ layer; 3 (squares) – TFLN with Ti layer; 4 (crosses) – TFLN with TiO$_2$ layer; 5 (triangles) – SOI with Ti layer; 6 (rhombuses) – SOI with TiO$_2$ layer.

The waveguide modes were simulated using the COMSOL Multiphysics software package. The influence of metallic Ti and dielectric oxidized Ti in the form of amorphous titanium dioxide (TiO$_2$, $n_{TiO2}' = 2.31$ [35]) films deposited on the upper surface of the waveguide was investigated for the film thicknesses ranging from 2.5 to 100 nm. Fig. 3a presents propagation loss versus Ti film thickness. The Ti film induces the propagation loss that has a broad maximum for a maximal overlap between the conducting Ti layer and the waveguide mode field. As the loss induced by the Ti film grows, the effective refractive index decreases (Fig 3b), and the light intensity maximum shifts slightly into the substrate depth. The $TiO_2$ layer does not produce any loss but increases effective refractive indexes (Fig. 3b).

Comparison of the results obtained for different material platforms leads to the conclusion that the influence of Ti metal overlay on the waveguide mode properties is higher for the waveguides with a larger numerical aperture since they have a greater overlap between the cover layer and the waveguide mode field.

## 3. Titanium laser oxidation

Laser oxidation of a thin Ti film was earlier used for the formation of micro- and nano-structures of diffractive elements [32]. The possibility of repeatable recording of a titanium dioxide line narrower than a diffraction limited focused laser spot [33,34] was demonstrated. Thus, this technique looks very promising for photonic circuit trimming where a high resolution is required, especially for waveguides with a high numerical aperture, such as a SOI waveguide.

The physical mechanism of laser oxidation is as follows. Laser radiation is absorbed by the titanium film and causes local heating, which depends on the laser power, spot size and thermal conductivity. The oxidation process is described by the Cabrera-Mott theory [48] as a migration of oxygen ions in the titanium oxide layer caused by an internal electric field (Mott potential) formed due to the excitation of titanium atoms. The excitation is highly dependent on local temperature. To provide local heating to a given temperature and start the oxidation process, a higher optical power is required for thicker films with a higher heat transfer coefficient.

As oxidation progresses, two counteracting processes occur. On the one hand, as the oxide thickness grows, the internal field decreases, and the oxidation process slows down. On the other hand, the metal thickness and the heat transfer coefficient decrease, which leads to an increase in local temperature. Stable oxidation is observed as a balance between these processes.

After through-thickness oxidation of the transparent substrate occurs, the absorption of laser radiation sharply decreases, heating stops, and the temperature drops, thus leading oxidation to a halt.

For extremely thin titanium film of a few nanometers thick, an evaluated local temperature lower than 200 °C is needed for oxidation process initialization [49]. The temperature practically does not change during oxidation and rapidly drops after through-thickness oxidation. These are fairly mild terms, compatible with most of PIC material platforms.

A set of samples was prepared for experimental investigations of the Ti film laser oxidation. Fused silica and *x*-cut congruent single crystal $LiNbO_3$ were used as substrates. Y propagated Ti:$LiNbO_3$ single-mode (at 1550 nm) optical waveguides were produced on $LiNbO_3$ substrates. Ti films of different thicknesses (5, 10, 15, 25, and 100 nm) were deposited on the substrate by DC magnetron sputtering.

Figure 4 presents the setup that was used for both titanium films laser oxidation and experimental investigation and PTF of PIC demonstration. Simple semiconductor laser pumping of continuous-wave from erbium doped fiber-optic amplifier was used. The sample substrates had low absorption at laser wavelength of 978 nm. Therefore, laser irradiation mostly affected the Ti films and did not affect the substrate materials. Note that other substrate material platforms require a different wavelength in order to oxidize Ti without affecting the substrate. For instance, SOI waveguides have significant absorption at 978 nm, so a longer wavelength should be used. Conveniently, the laser thermal Ti excitation used to drive the oxidation process can be realized over a wide wavelength range.

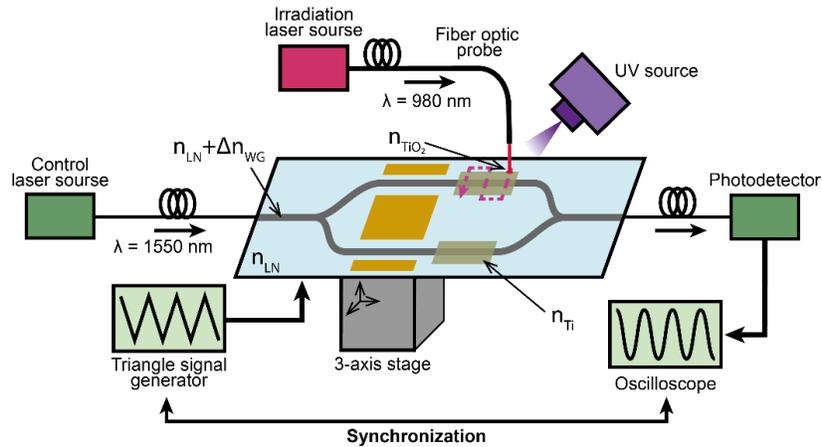
Fig. 4. Experimental setup for local laser oxidation of a titanium film.

The output single-mode fiber (PM980-HP) of the laser was used as a probe for local laser exposure. The fiber tip was positioned close to a sample surface (at a height of about 50 μm). The laser spot was about 10 μm in diameter, and the maximum peak intensity was about 1 kW/mm$^2$. The sample was attached to a precision 3-axis translation stage that provided the laser spot movement along a given trajectory on the sample surface. The scanning speed of the fiber probe was 1 mm/s. As in the previous study devoted to laser Ti oxidation [32,33], all the processes were carried out in air, that is without a specific gas atmosphere. To suppress the effects inherent in lithium niobate related to space charge forming, in particular, the photorefractive effect, additional homogeneous UV illumination was used.

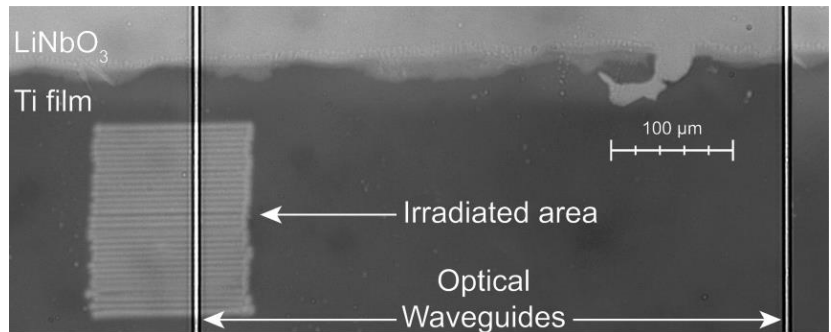
Fig. 5. Transparent traces on the metallic Ti film after laser irradiation.

Initially, the change in the morphology of Ti films under laser irradiation was investigated. The stripes from the straight single path of the fiber probe and rectangular areas from the meander-like multi path movement of the fiber probe were formed by laser irradiation of the sample surface (Fig 5) [26]. A meander-like path with a half period of 5 μm was used to provide path overlapping and oxidized rectangular areas continuity. The laser action manifested itself in the form of transparent traces on the metallic Ti film. Swelling was seen in the AFM images in the laser irradiation area. The examples of height profiles of oxidation of a 5 nm thick Ti film on a LiNbO$_3$ substrate are shown in Fig. 6 [26]. The swelling height was about 10 nm. Estimated by change in molar mass from 47.867 g/mol (Ti) to 79.866 g/mol (TiO$_2$) and change in film density from 4.5 g/cm3 (Ti) to 3.1 g/cm3 (thin-film amorphous TiO$_2$ film deposited by reactive evaporation [50]) gave thickness increase about 7 nm. This discrepancy could be due to the uncertainty in thickness of metallic Ti film, which was set by the conditions of DC magnetron sputtering, as well as to a more complex film composition after laser induced oxidization than that of TiO$_2$. One can see a periodic contour at the vertices of rectangular areas. This is attributed to the meander-like path of the fiber probe scanning. The periodical pattern could contribute to scattering and addition loss for guided light that were not taken into account in the theoretical model. However, as experimental study showed, the contribution of scattering at a sufficiently small distance used for PFC had been much smaller than the loss on metallic film and insignificant.

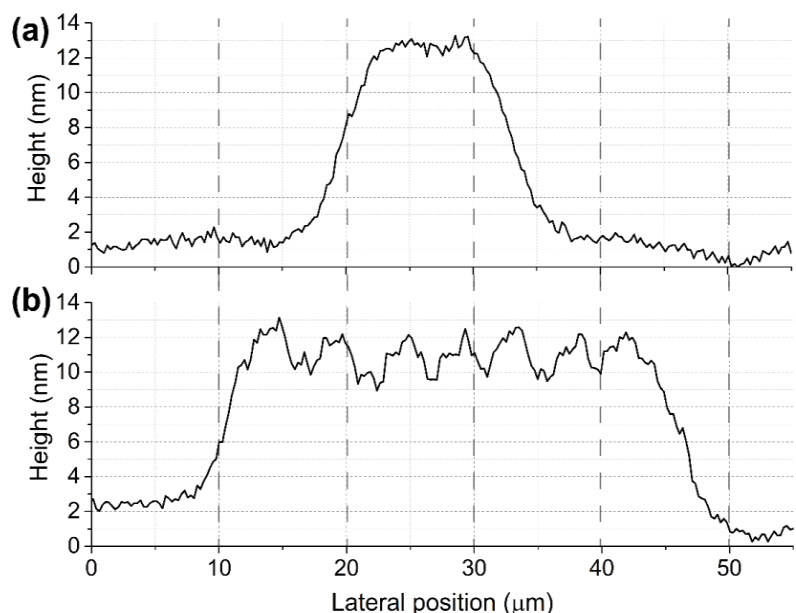

Fig. 6. The height profiles of the 5 nm thick Ti film areas irradiated with (a) single motions of the probe and (b) meandering motions of the probe.

Then the composition of the films was studied by X-ray photoelectron spectroscopy (XPS). The lines related to titanium dioxide ($TiO_2$) in the XPS spectra of the irradiated areas indicated that laser-induced oxidation of titanium took place. The carbon line was attributed to surface contamination. Plasma etching was used to study the change in film composition with the depth of the oxide film. The XPS spectra of very thin films (5, 10 and 15 nm) did not alter with depth, and the titanium line (Ti 2p) disappeared after the film was completely removed from the substrates (Fig. 7a). There was evidence of homogeneous oxidation and an additional confirmation of a very weak effect on the substrate material. The results can be well described by the physical model, in which the oxidation of very thin Ti films starts at a relatively low temperature (< 200 °C) and stops quickly after through oxidation.

Irradiated 100 nm titanium film XPS spectra at different ion etching depth are shown in Fig. 7. Curve (1) reveals only $TiO_2$ after the removal of contaminations from the top of the film (etching the 4 nm layer off the film surface). Decomposition of Ti 2p doublet (Fig. 7c) demonstrates three different valence states of Ti after the removal of 8 nm layer from the film (curve (2) in Fig. 7b). Curve (3) in Fig. 7b corresponds to metallic Ti. It was obtained after 170 nm etching, as the film thickness was increased after laser treatment. The XPS spectrum after the film complete removing from the $LiNbO_3$ substrate (etching off 370 nm film) contains some titanium traces possibly associated with Ti-indiffusion into the substrate. Apparently, after a certain threshold thickness of the Ti film, the oxidation process began to affect the substrate material.

Inhomogeneities and defects in the oxidized 20 nm Ti film were also observed in scanning electron microscope (SEM) images (Fig. 8). In accordance to the physical model, the film was locally heated to a higher temperature that could potentially lead to film defects and indiffusion into the substrate.

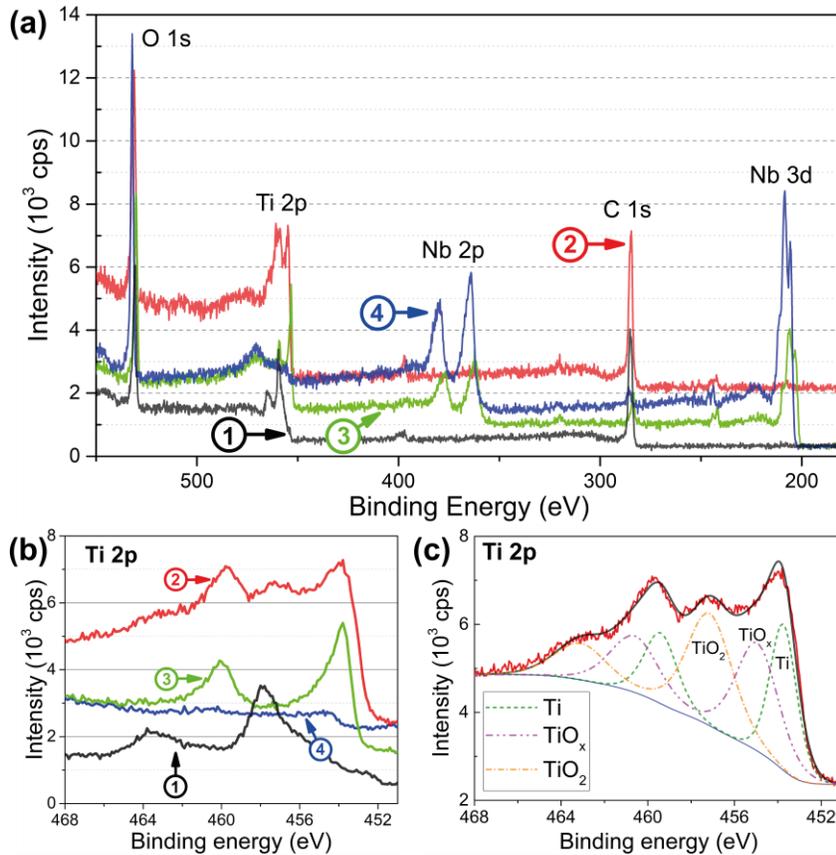

Fig. 7. (a) XPS spectra of irradiated 100 nm titanium film at different ion etching depth: the black line (1) – 4 nm, the red line (2) – 8 nm, the green line (3) – 170 nm, the blue line (4) – 370 nm; (b) Ti 2p doublet XPS spectra; (c) decomposition of Ti 2p doublet XPS spectra after 8 nm etching.

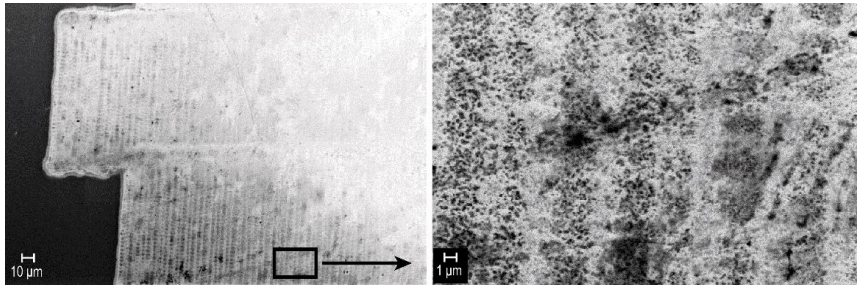

Fig. 8. SEM images with inhomogeneities and defects in the 20 nm titanium film after laser irradiation and through-thickness oxidation.

Despite a crystal phase of oxidized films has not been studied, it should be amorphous for thin film (< 20 nm) formed at low temperatures (< 200 °C). High-temperature annealing is required to obtain the anatase and rutile phases. In addition, the influence of the crystal phase on the waveguide optical properties is very weak and negligible in comparison with transition from metallic Ti to dielectric TiO2 in any crystal phase for very small film thickness (several nanometers). For thicker films, the formation of a mixture of polymorphs is highly likely.

Finally the effect of the film on Ti:LiNbO$_3$ waveguides was studied. As predicted by the theoretical analysis in Section 2, the titanium film oxidation should reduce the optical loss. To investigate it, the rectangular areas of laser oxidation were intentionally produced over

Ti:LiNbO$_3$ waveguides on the LiNbO$_3$ substrates. Input and output single mode optical fibers were attached to optical waveguides for real-time optical loss monitoring. The change in waveguide loss per unit length during laser oxidation was measured by scanning the meander path of the fiber probe across the waveguide channels. The theoretical prediction was confirmed by the experimental dependence of the effect of metallic Ti overlay on the insertion loss of Ti:LiNbO$_3$ waveguide (Fig. 3a).

## 4. Mach-Zehnder modulator trimming

As it was shown, the oxidation process is well explained in terms of Cabrera-Mott theory, and for a few nanometer-thick titanium film, an estimated local oxidation temperature for the oxidation does not exceed 200 °C. These are fairly mild conditions, compatible with most of the PIC material platforms. Besides, this oxidation mechanism, as was shown in [33, 34] could provide very high spatial resolution, which was required for PIC with a high refractive index difference (high numerical aperture (NA) and small element footprint). Our numerical estimation has shown that the effect of overlay layer is large enough for PIC post-fabrication, and it is more prominent for waveguides with a high NA, such as, for instance, SOI-based waveguides.

The proposed technique of Ti film laser oxidation was used to demonstrate experimentally the PFT of the LiNbO$_3$ Mach-Zehnder modulator (MZM). This material platform uses LiNbO$_3$ crystalline substrates that are transparent in a wide spectral range. A typical MFD of waveguides is about 10 μm, so a PFC does not require a very high spatial resolution, which simplifies the technical implementation. However, the influence of the overlay layer is also less for Ti:LiNbO$_3$ waveguides, and various effects inherent for lithium niobate (such as piezo, pyro, photorefractive and other ones) hinder PFT monitoring.

Samples of the integrated optical MZM were fabricated on the *x*-cut congruent single crystal LiNbO$_3$ substrate (Fig. 4) by titanium thermal in-diffusion. Light propagated along the *Y* crystallographic axis. Two single mode fiber optic pigtails (5 mm × 50 mm × 1 mm) were attached to the chip to input and output an optical signal using the end-fire coupling technique [51]. The total optical loss, including coupling fiber-to-waveguide and waveguide-to-fiber loss, was about 4.2 dB and they slightly varied from one sample to another. The modulator half-wave voltage was $U_\pi = 3.1$ V, and the extinction ratio was in the range of 20–30 dB, which are typical parameters for commercially available devices.

The 5 nm thick Ti was deposited over the waveguides in the area of two parallel arms clear of electrodes. It covered both waveguides and was 2 mm in length along the light propagation direction. The additional losses attributed to the Ti film were about 2 dB, as was predicted by the theoretical analysis. A rather long length of the area covered with Ti was chosen advisedly for versatility in an iterative MZM adjusting. According to theoretical estimations, this length provides about 2 dB tuning range of optical losses and 0.1 rad tuning range of phase difference.

MZM adjusting was performed on the same setup that was used for the experimental investigation of laser Ti film oxidation process (Fig 4). On-line monitoring of the modulator transfer function was carried out during the trimming process. A fiber pigtailed distributed feedback laser diode (DFB LD) having wavelength of 1550 nm and output power of 13 dBm was used as a signal light source. A triangular signal with a peak-to-peak voltage of 30 V, which was 10 times higher than the modulator half-wave voltage ($U_\pi = 3.1$ V), was applied to the modulator from a signal generator. Modulated optical signal at the modulator output was detected by a p-i-n photodiode (PD) with a transimpedance amplifier with tunable gain and was displayed on a digital oscilloscope. Figure 9a shows a typical cosine transmission characteristic. A homogenous illumination of the samples by low intensity (~ 1 mW/mm$^2$) incoherent UV light was additionally applied to suppress the photorefractive response of LiNbO$_3$ to laser irradiation [52] and ensure that only the Ti film oxidation affected the modulator.

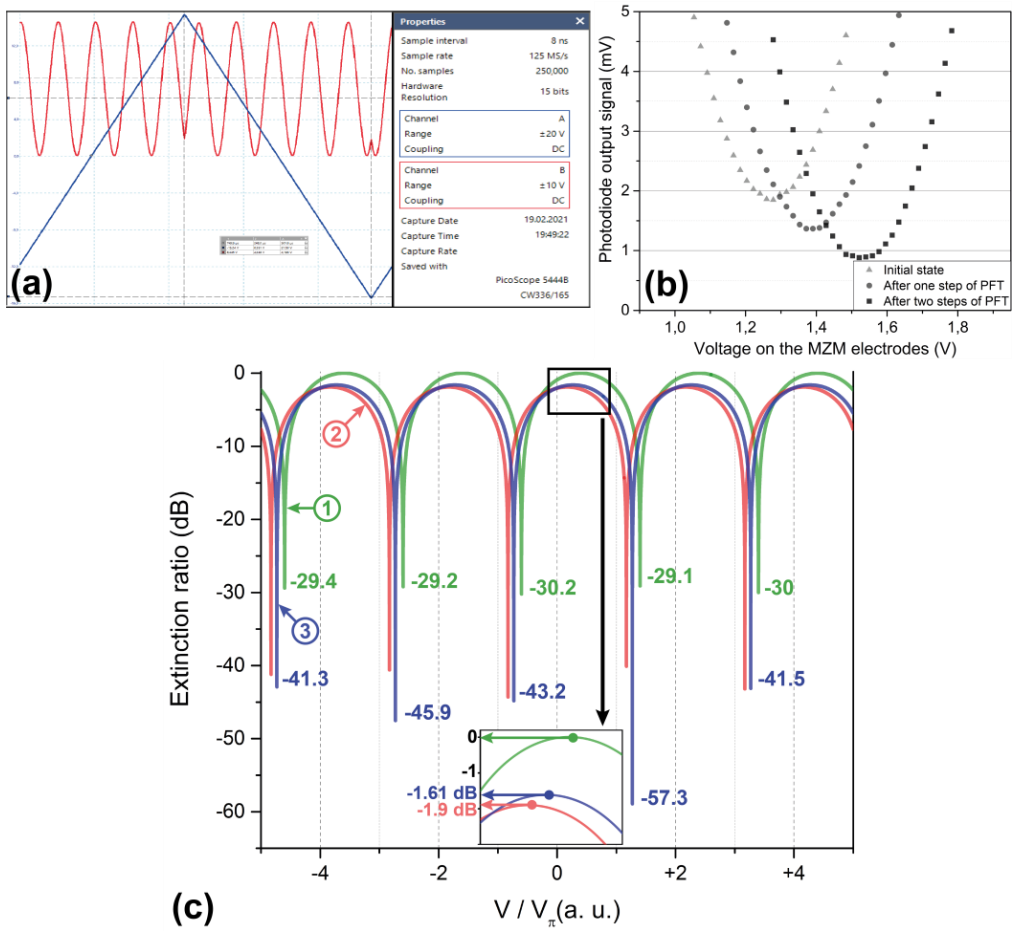

Fig. 9. (a) on-line monitoring of the cosine modulation response of the MZM to an input triangular signal; (b) refined measurements of transmission minimum after sequential laser oxidation of the Ti film with a step of 100 μm along optical waveguide; (c) enhancement of extinction ratio in the MZM: an initial response of the MZM (1), the response after the Ti film deposition before PFT (2) and after PFT (3).

Two types of the MZM trimming were demonstrated. Firstly, an improvement in the extinction ratio was achieved. The initial extinction ratio of the samples was in the range of 20 – 30 dB. The reduction in the optical loss caused by the Ti film oxidation was used for balancing the light intensity in the Mach-Zehnder interferometer arms by sequential illumination of different arms of the waveguide interferometer. On-line monitoring of the modulation contrast in the cosine modulation electrooptic response of the MZM (transfer function) was used for a proper choice of the waveguide arm for local laser oxidation and on-line termination of the fiber optic probe scanning (oxidation termination).

Figure 9c illustrates the change in the extinction ratio depicted in a logarithmic scale. The maxima of the transmission characteristics and positions of minima were defined during on-line monitoring of MZM transmission characteristic. The values at the minima of the transmission characteristics were refined using additional measurements. To reliably detection a very weak signal at the minima, we used the maximum gain of transimpedance amplifier and a long averaging time (1 s). The voltage corresponded to a minimum of the modulator transfer function defined in these coarse measurements with triangular signal was set on the MZM electrodes. Then the voltage was fine tuned to minimum signal on PD output manually (Fig. 9b). Note that position of the minimum shifted, and an additional fine tuning was required after each successive PFT. An increase in the extinction ratio by 27 dB from 30 dB up to 57 dB was achieved (Fig. 9c). Note that the extinction differs in different minima. We explain these variations as MZM imperfections that cannot be compensated by waveguide propagation loss balancing, in particular, the interaction with light leaked from waveguide into substrate at fiber-to-chip coupling point and on the Y-branches.

The decrease in the MZM optical loss after PFT was about 0.29 dB and attributed to shortening of metallic titanium film. The difference of the film length in the arms of MZM after PFT was about 350 μm.

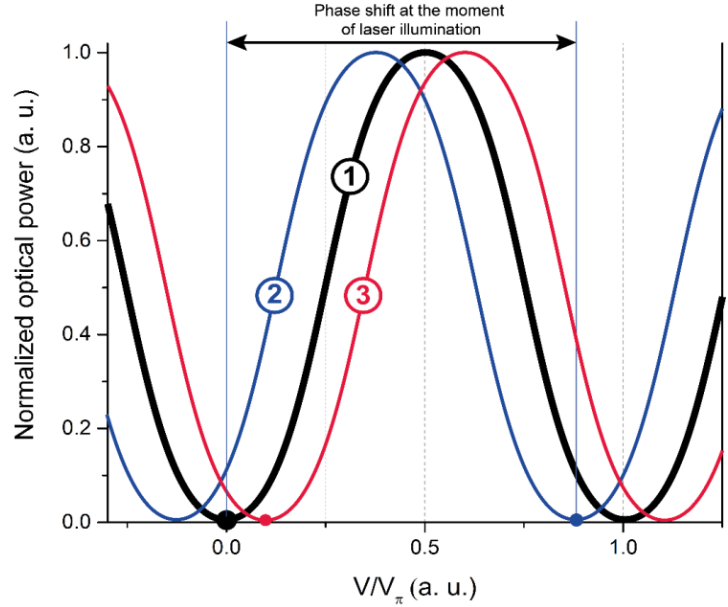

Fig. 10. Cosine transfer function minimum shift. The black line (1) indicates the position before laser irradiation; the blue line (2) indicates the phase shift just after irradiation of the Ti-film with laser (single pass of the fiber probe across the waveguide); the red line (3) is the stable relative cumulative phase shift after 40 successive passes of the zigzag (220 μm) across the waveguide in one of the MZM arms. When measuring the phase shift, only the positions of the transmission characteristic minimum were tracing. Cosine transmission characteristics are shown for easier understanding.

The change in the cosine transfer function contrast was accompanied by the electrooptic response shift. The shift of the transfer function minimum shown on the Fig. 10 is a cumulative effect of successive fiber optic probe scanning. This effect could be used for tuning the MZM operating point and was studied in more details. The position of the transmission characteristic minimum was tracing during phase shift measurements. Electronic system with feedback [53] was used for locking the minimum. Characteristic time constant of the system was 12 ms and define time resolution of the phase shift monitoring.

A significant transient phase shift (about π) was observed during several seconds even for a single fiber optic probe pass across an arm of the waveguide interferometer. The initial significant phase shift occurred when the fiber probe was outside the optical waveguide. It increased as the probe moved toward the waveguide and then relaxed to a fixed value when the probe crossed the waveguide and moved away. Supposedly, this behavior could be partially explained as a response on local heating during Ti oxidation (a combination of thermal refractive index changes and pyroelectric effect in $LiNbO_3$).

The fixed value of the phase shift is maintained after transient process. It was very stable and did not change at a periodic monitoring for several months. Gradual tuning of the position of the cosine transfer function minimum was demonstrated by step-by-step cumulative tuning pass by pass (Fig. 10).

A pause of ten seconds between successive pass ensured the relaxation of the transient phase shift. The tuning step was about $4.07 \times 10^{-3}$ V/V$_\pi$ (0.0128 rad). It corresponded to a single pass of the fiber probe across the waveguide and was defined by spatial resolution (laser spot size), which was about 10 μm. A fixed phase shift was observed only at laser irradiation of sample areas coated with a Ti film, and in contrast to the reversible photorefractive phase shift [54], which was also observed at irradiation of waveguides without a Ti film, could not be suppressed by homogeneous UV illumination of the sample.

It should be also pointed up that the fixed phase shift was higher than predicted by theory, which was attributed to the effective refractive index change caused by Ti oxidation. Further research is needed to explain this result. Possibly, the mechanical stress caused by the TiO$_2$ overlayer plays a role.

Nevertheless, once again we emphasize the stability of the MZM characteristics after adjustment, which is a required feature of a permanent PPT. The extinction ratio slightly decreased (by 3 dB) during the first week, which can be attributed to a partial suppression of the photorefractive effect and a partial relaxation of the mechanical stress caused by laser oxidation. Thereafter, the maximum extinction ratio (57 dB) remained stable at least 4 months of periodical monitoring. The long-term stability of the MZM operating point was difficult to trace because of the strong influence of ambient conditions, primarily temperature.

## 5. Conclusion

In conclusion, a novel technique for post-fabrication trimming of photonic integrated circuits has been proposed. The technique was based on local laser oxidation of a thin Ti film. The investigation of the oxidation process has shown that this process is well explained in the framework of the Cabrera-Mott theory for negligible laser absorption in a waveguide and in a substrate. For a few nanometers thick titanium film, the estimated local oxidation temperature does not exceed 200 °C. Thus, the considered technique has very little effect on the waveguide and substrate materials and can be classified as a technique of cladding modification. It can be applied to most dielectric optical waveguide material platforms by choosing the proper laser wavelength, and it is CMOS compatible. Laser oxidation of Ti films effectively regulates both the propagation loss and the effective refractive index of dielectric optical waveguides. Thus, a versatile PFT of PICs can be realized. The nonlinear mechanism of laser oxidation related to the threshold laser intensity for oxidation in the framework of the Cabrera-Mott theory can provide lateral self-limitation and high spatial resolution not restricted by a diffraction-limited size of the laser spot [33,34], which looks promising for PFT of elements with a small footprint.

The efficiency of this technique was demonstrated as PFT of Ti:LiNbO$_3$ MZM. Lithium niobate is a rather complicated material for permanent PFT. It has many effects, such as piezo, pyro, photorefractive and other ones, which, on the one hand, provide an additional tuning mechanism, but, on the other hand, are volatile and difficult to control. We used homogeneous UV illumination to suppress the effects related to space charge forming, in particular, the photorefractive effect. Fine tuning of the operating point and fixed extinction ratio enhancement by 27 dB (up to 57 dB) has been demonstrated. No long-term trimming relaxation was observed during several months of periodic monitoring. These results may be interesting for some applications such as quantum telecommunications and precise optical measurements.

It's worth remarking that the efficiency of PFT according to the proposed technique should be higher for waveguides with a higher refractive index contrast (NA) and a smaller MFD, such as those fabricated on the basis of thin-film lithium niobate (TFLN), silicon nitride, or silicon on insulator (SOI). Thus, for these material platforms more efficient photonic device trimming in a wider range of parameters is expected. However, it will be necessary to select the optimal laser wavelengths for oxidation and more precise laser beam control, focusing and alignment.

**Funding.** Russian Science Foundation (19-19-00511).

**Acknowledgments.** This work was supported by the Russian Science Foundation project 19-19-00511.

**Disclosures.** The authors declare that there are no conflicts of interest.

**Data Availability.** Data underlying the results presented in this paper are not publicly available at this time but may be obtained from the authors upon reasonable request.